\documentclass[journal]{IEEEtran}

\usepackage{cite}
\usepackage{amsmath, amsfonts}
\usepackage{xcolor}
\usepackage{siunitx}
\usepackage{graphicx}
\usepackage{epstopdf}
\usepackage{booktabs}
\usepackage{tabularx,ragged2e,booktabs,caption}
\usepackage{float}
\usepackage{multirow}
\usepackage{mathtools}

\begin{document}

\onecolumn
\copyright 2020 IEEE. Personal use of this material is permitted. Permission from IEEE must be obtained for all other uses, in any current or future media, including reprinting/republishing this material for advertising or promotional purposes, creating new collective works, for resale or redistribution to servers or lists, or reuse of any copyrighted component of this work in other works
\twocolumn

\newpage

\title{A Millimeter-scale Single Charged Particle Dosimeter for Cancer Radiotherapy}

\author{Kyoungtae Lee, %~\IEEEmembership{Student Member,~IEEE,}
        Jessica Scholey,
        Eric B. Norman, 
        Inder K. Daftari, 
        Kavita K. Mishra, 
        Bruce A. Faddegon, 
        Michel M. Maharbiz,%$^\dagger$,~\IEEEmembership{Senior Member,~IEEE,}
        ~and~Mekhail Anwar%$^\dagger$,~\IEEEmembership{Member,~IEEE}% <-this % stops a space
%\thanks{$^\dagger$ M. M. Maharbiz and M. Anwar contribute equally to this work as senior authors.}
\thanks{M. M. Maharbiz and M. Anwar contribute equally to this work as senior authors.}
%\thanks{K. Lee is with the Department of Electrical Engineering and Computer Sciences, University of California, Berkeley, CA 94720 USA (e-mail : ktlee@berkeley.edu).}% <-this % stops a space
%\thanks{J. Scholey, I. K. Daftari, K. K. Mishra, B. A. Faddegon, and M. Anwar are with the Department of Radiation Oncology, University of California, San Francisco, CA 94143 USA.}% <-this % stops a space
%\thanks{E. B. Norman is with the Department of Nuclear Engineering, University of California, Berkeley, CA 94720 USA.}%
%\thanks{M. M. Maharbiz is with the Department of Electrical Engineering and Computer Sciences, University of California, Berkeley, CA 94720 USA, and also with Chan-Zuckerberg Biohub, San Francisco, CA 94158 USA.}%
}

%\markboth{IEEE Journal of Solid-state Circuits}%
\markboth{\copyright ~2020 IEEE}
{Shell \MakeLowercase{\textit{et al.}}: Bare Demo of IEEEtran.cls for IEEE Journals}

\maketitle

%Var(a*X) = a^2 Var(X)
%Var(ln(X)) = Var(X)/mean(X)^2
%Var(a*ln(X)) = a^2 Var(X)/mean(X)^2
%Var(a*ln(c/X)) = a^2(Var(X)/mean(X)^2 )
%Var(a*ln( c/(d-X)) = a^2(Var(y)/mean(y)^2) = a^2(Var(x)/(d-mean(x))^2) = a^2(Var(x)/d^2)

\begin{abstract}
This paper presents a millimeter-scale CMOS 64$\times$64 single charged particle radiation detector system for external beam cancer radiotherapy. A 1$\times$1 $\mathbf{\mu m^2}$ diode measures energy deposition by a single charged particle in the depletion region, and the array design provides a large detection area of 512$\times$512 $\mathbf{\mu m^2}$. Instead of sensing the voltage drop caused by radiation, the proposed system measures the pulse width, i.e., the time it takes for the voltage to return to its baseline. This obviates the need for using power-hungry and large analog-to-digital converters. A prototype ASIC is fabricated in TSMC 65 nm LP CMOS process and consumes the average static power of 0.535 mW under 1.2 V analog and digital power supply. The functionality of the whole system is successfully verified in a clinical 67.5 MeV proton beam setting. To our' knowledge, this is the first work to demonstrate single charged particle detection for implantable \emph{in-vivo} dosimetry.    
 
\end{abstract}

%\begin{IEEEkeywords}
%IEEE, IEEEtran, journal, \LaTeX, paper, template.
%\end{IEEEkeywords}

\IEEEpeerreviewmaketitle

\section{Introduction}

\IEEEPARstart{M}{ore} than half of cancer patients are treated with ionizing radiation, where the fundamental goal is to deposit sufficient energy (dose) to destroy the tumor cells and stop their proliferation. A key challenge in radiotherapy is to target the tumor while imparting minimal damage to surrounding normal tissues. Commonly-used \emph{external beam radiotherapy} (EBRT) employs x-ray photons to deliver a radiation dose to the tumor. This well-established method encounters several difficulties: 1) X-rays pass through the whole body, leaving unwanted dose in healthy tissues. This can be critical in pediatric cancer, for example, where secondary malignancy results from peripheral dose; 2) The dose from x-rays is highest near the surface, dropping a few percent per centimeter with depth. 

\begin{figure}[!t]
\centering
  \includegraphics[width=250 pt]{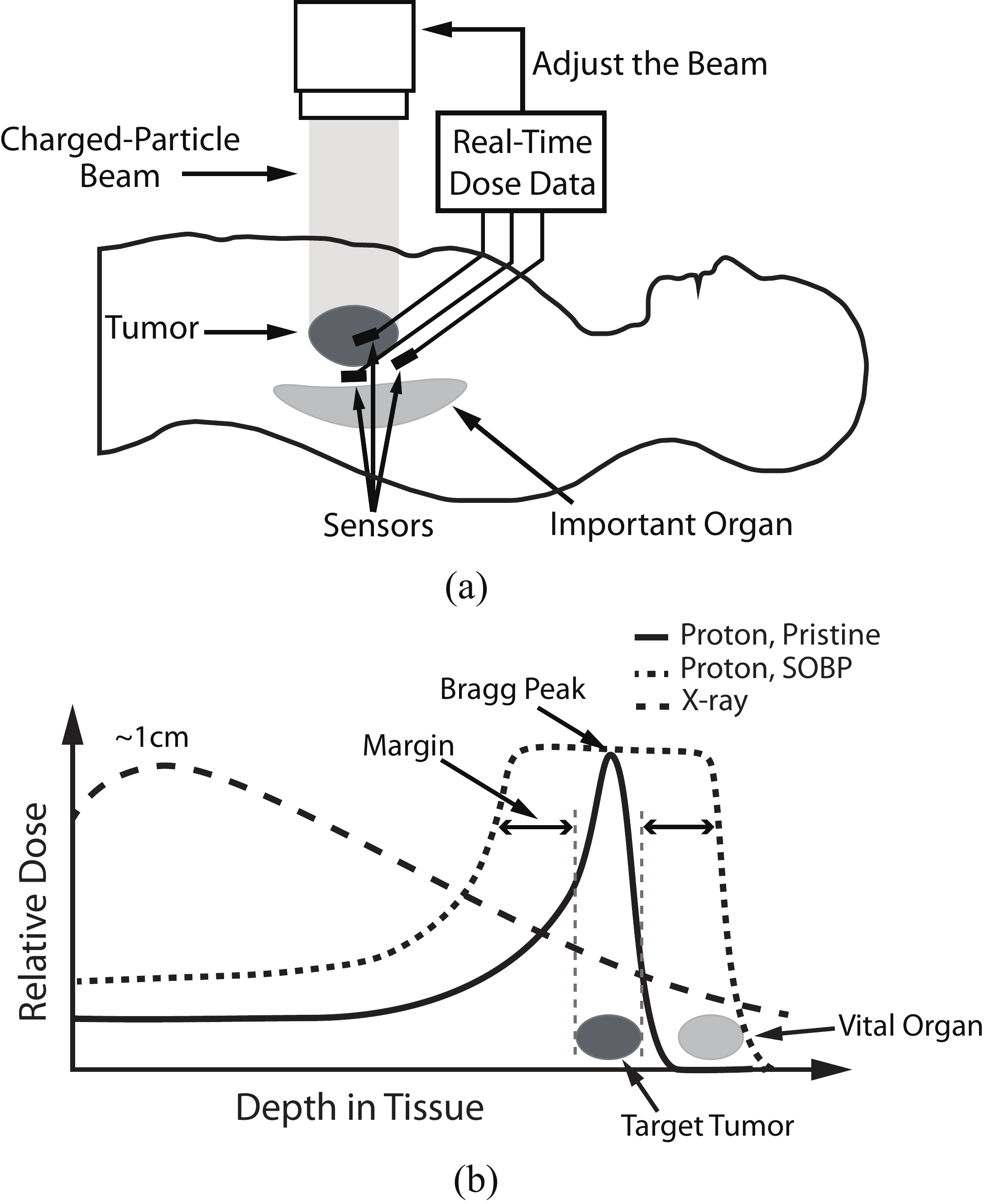}
  \caption{Illustrations of (1) cancer treatment with \emph{in-vivo} dosimetry and (b) depth-dose curve for X-ray, proton (pristine), and proton (SOBP).}
  \label{fig:fig1}
\end{figure}

Due to these issues, \emph{charged particle radiotherapy} has advantages over x-rays. Unlike photons, charged particles (such as protons and carbon ions) deposit the highest dose in a specific location at the end of their range (the Bragg peak), theoretically allowing dose to be delivered with higher precision and with less peripheral dose than with x-rays (See Fig. \ref{fig:fig1} (b)).  

Despite the advantages of charged particle therapy, a current limitation is knowing the exact location of the Bragg peak (range uncertainty) which is caused by a number of factors. First, patient movement such as respiratory motion shifts the Bragg peak. Second, charged particle interactions occurring within the body depend heavily on tissue atomic properties, which are difficult to determine accurately \cite{protonphysics}. Lastly, day-to-day anatomical variations may make predictions inaccurate. 

The current clinical practice is to mitigate the range uncertainty by using a patient-specific maps of estimated particle stopping power derived from CT image to predict the location of the Bragg peak. This method is rooted in the stoichiometric method \cite{stoi}, which provides a parametrized fit of CT Hounsfield Units to the stopping power ratio of material. However, this comes at a price in range uncertainty due to potential errors in converting the X-ray attenuation coefficient to proton stopping power as well as uncertainties in the patient CT image. In addition, stoichiometric calibration cannot solve daily anatomical variations and patient movement issues. As a result, the typical range uncertainty is about 2.5 $\%$ of the total range. For example, if the Bragg peak is predicted to fall $\SI{100}{\mm}$ inside a patient body, a range uncertainty of $\SI{2.5}{mm}$ will significantly impact the precision of the dose delivery. Given this, it is common in clinics to widen the Bragg peak to cover the full target volume, and then add treatment margins to ensure the target is covered with prescription dose, resulting in increased dose to normal tissue. In addition, sub-optimal beam arrangements may be selected to avoid delivering dose to a critical organ just distal to where the proton beam stops. An example of a spread-out Bragg peak (SOBP) and the additional margins added to account for this range uncertainty is illustrated in Fig. \ref{fig:fig1} (b)).

Real-time \emph{in-vivo dosimetry (IVD)} ameliorates uncertainty by measuring the dose delivered in the body, potentially leading to more effective and safer closed-loop treatments (Fig. \ref{fig:fig1} (a)). Clinically viable IVDs have several important constraints. They must be millimeter scale for implantation through a standard core-biopsy needle; consume very small amounts of power; have single-particle sensitivity; be capable of real-time measurement of energy deposition; and be suitable for bio-compatible chronic implantation (usually 1-8 weeks) with appropriate medical-grade packaging. These requirements strongly drive the need for a CMOS platform capable of compact integration of low-power sensors and readout circuitry.  

While existing approaches have made progress towards miniaturized IVD, no previous work has satisfied all requirements \cite{MOS1, MOS2, RL, FG, Diode, PSD}. Single MOSFET dosimeters have been the most widely used, as they can be easily fabricated in a small size \cite{MOS1, MOS2}. Integrated damage by radiation in the SiO$_2$ layer of the MOSFET decreases the threshold voltage linearly. However, the lifetime is finite and it lacks single particle sensitivity due to the cumulative nature of the radiation induced damages. Plastic scintillator, thermo-luminescent, or radio-luminescent dosimeters detect light intensity when a radio-sensitive material is exposed to radiation \cite{RL, PSD}. However, they measure only cumulative dose, cannot provide real-time data, and require bulky optical equipment to measure light that precludes implantation. Floating gate dosimeters measure the current change when charges are trapped in the floating gate by radiation \cite{FG}, but they lack single particle sensitivity. 

%key metric to biologic damage is energy deposited with the particle, but nonlinear. therefore, measuring the cumulative damage loses info. so, single particle detection is a key feature. and discuss the paper.  
Most importantly, conventional dosimeters measure average dose, and ignore a critical phenomenon: for a given dose, a single high linear-energy-transfer (LET, energy deposition per unit length) particle has a significantly different biological effect on tissue than that of several low LET particles \cite{BioEffect, BioEffect2}. The key metric to the biological effect is the energy deposition by each particle. Because the biological effect versus energy deposition is non-linear, the cumulative damage does not represent the true biological effect by radiation. For example, the normalized average number of lethal lesions in a HF19 human diploid fibroblasts cell produced by a single $4$, $50$, and $70$ $\SI{}{keV}/\SI{}{\mu m}$ LET alpha-particle is approximately $1$, $34.4$, and $65.6$, respectively \cite{BioEffect}. The biological effect increases more rapidly than the energy deposition. The biological effect plateaus after $\SI{100}{keV}/\SI{}{\mu m}$. Due to this non-linearity relationship between the energy deposition and the biological effect, single particle detection will be a key feature for next-generation IVDs, enabling analysis of the true biological effect by radiation.  

In this work, we solve these challenges by introducing a 64$\times$64 millimeter scale single charged particle CMOS dosimeter, compatible with \emph{in-vivo} implantation for EBRT. To the best of our' knowledge, the proposed system is the first work to enable single charged particle detection using only conventional CMOS chip fabrication process. 

\section{Theory of operation}
This section describes how protons interact and deposit energy in matter, and how the deposited energy relates to the biological effect. The expected signal measured by a diode is analyzed. Finally, the acquisition of a pulse width (as opposed to a voltage level measurement) is discussed. 

\subsection{Proton interaction with matter}
With the clinically-relevant energy range, protons deposit energy when passing through matter by three types of interactions: 1) Coulomb interactions with atomic electrons; 2) Coulomb interactions with atomic nuclei; and 3) nuclear reactions accompanied by creation of secondary particles (proton, neutron, electron, and gamma ray) \cite{protonphysics}. The first type is the most dominant type of interaction, where a proton ionizes matter, transferring part of its energy to electrons that deposit their energy in proximity to the point of ionization ($\sim$ \SI{1}{mm}). The second type alters the proton trajectory and contributes to proton scattering. The last type is the rarest. In the first type, LET describes the average amount of proton energy deposited per unit length, and is well-modeled by the Bethe-Bloch equation.

\begin{equation}
    %\frac{dE}{dx} = 4 \pi \rho N_A {r_e}^2 m_e c^2 \frac{Z}{A} \frac{z^2}{\beta^2} \left[ \ln \frac{2m_e c^2 \gamma^2 \beta^2}{I} - \beta^2 - \frac{\delta}{2}-\frac{C}{Z} \right]
    \frac{dE_{dep}}{dx} \propto \rho \frac{Z}{A} \frac{1}{\beta^2} \left[\ln{\frac{2m_ec^2\gamma^2 \beta^2}{I}-\beta^2-\frac{\delta}{2}-\frac{C}{Z} }   \right],
\label{eq:bethe}
\end{equation}

where $dE_{dep}/dx$ is the energy deposition per unit length, $\rho$ is the density of the absorbing material, Z is the atomic number of the absorbing material, A is the atomic weight of the absorbing material, $\beta = v/c$ where v is the velocity of the proton and c is the speed of light, $m_e$ is the electron mass, $\gamma=(1-\beta^2)^{-1/2}$, I is the average ionization potential of the absorbing material, $\delta$ is the density correction term, and $C$ is the shell correction term. Eq. \ref{eq:bethe} shows why it is challenging to predict the location of the Bragg peak, as the LET value heavily depends on the material property and proton energy.  

Dose (\si{\gray=\joule/\kilogram}) is widely used in clinical applications to quantify the radiation effect on tissue:

\begin{equation} 
    Dose = \sum_{i=1}^{N}{ \frac{E_{dep,i}}{m}} = \frac{\mathbb{E}[E_{dep}] \times N}{m},
\end{equation}
where $N$ is the number of protons, $E_{dep,i}$ is the energy deposition by each proton in the material, and $m$ is the mass of material where the energy deposition occurred. Dose is the sum of individual energy depositions per unit mass. However, the actual biological effect (e.g., the number of double strand breaks in the DNA or cell mortality rate) for particles with higher LET has a highly non-linear relationship with the $E_{dep,i}$ \cite{BioEffect, Book}. This means that dose alone is an insufficient measure to evaluate the true effect on tissue. We also need the LET; that is, the single particle detection sensitivity.

\begin{figure}[!t]
\centering
  \includegraphics[width=225 pt]{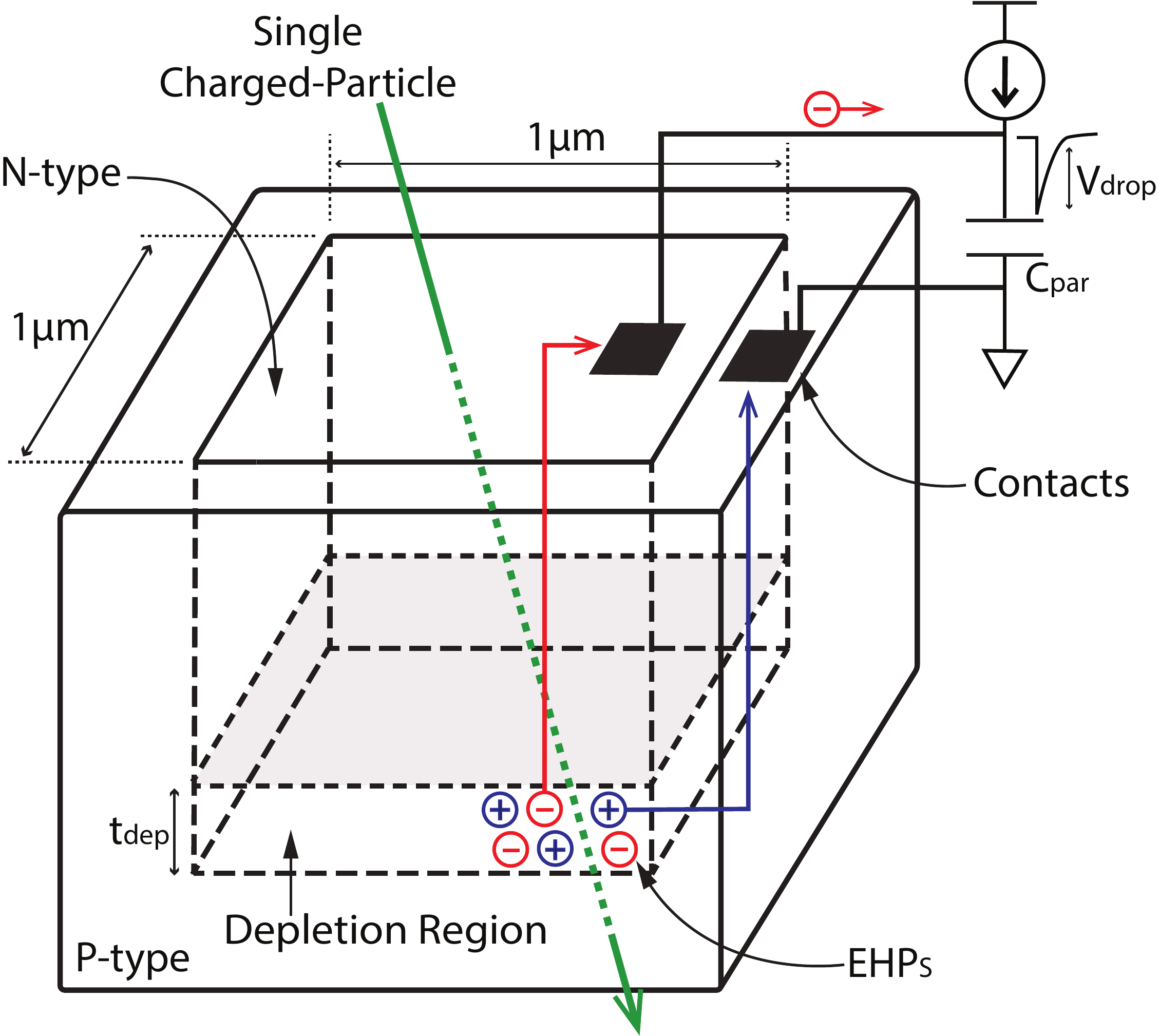}
  \caption{Illustration of diode sensing mechanism.}
  \label{fig:diode}
\end{figure}

\begin{figure}
  \includegraphics[width=252 pt]{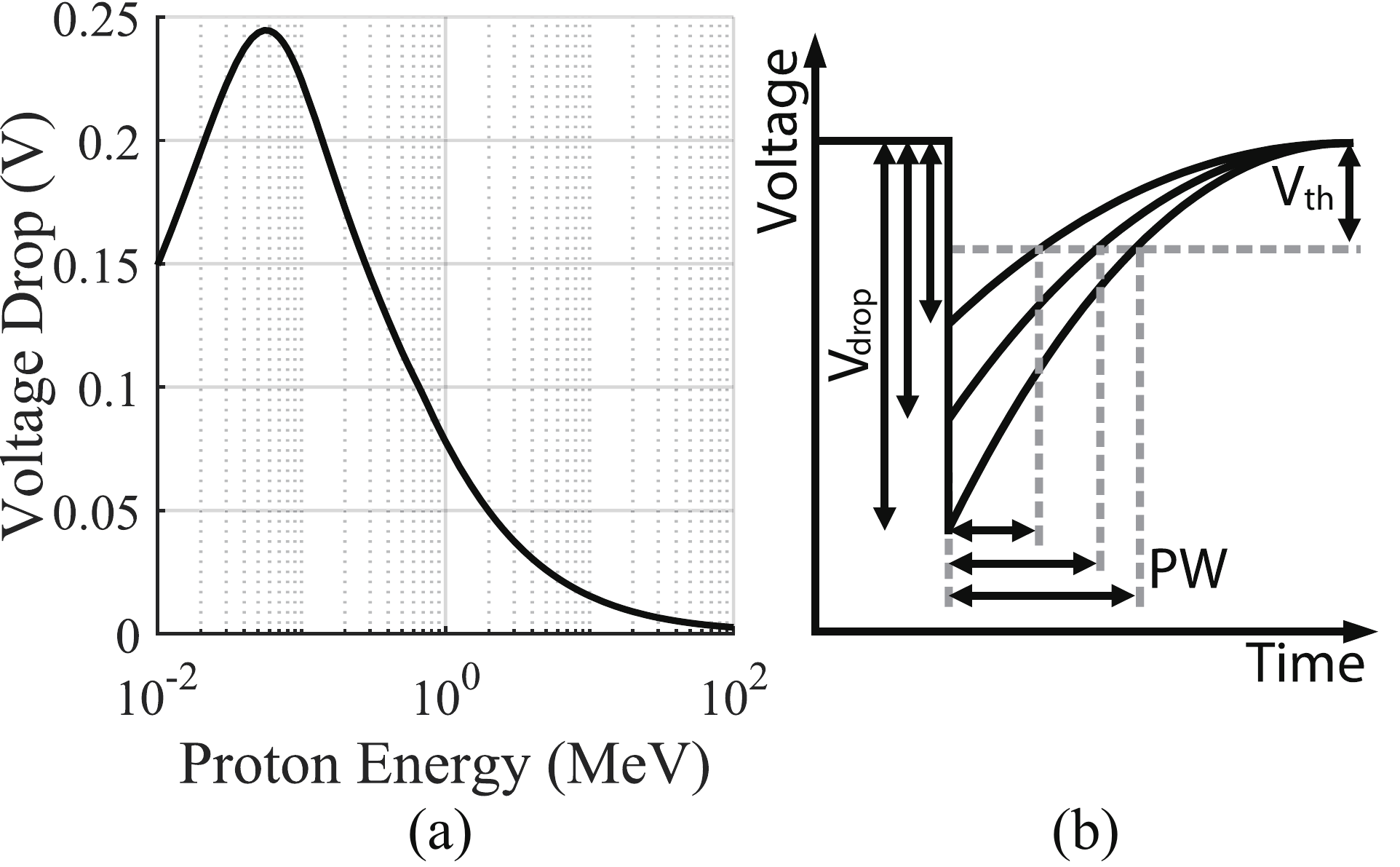}
  \caption{(a) Expected voltage drop at the diode sensing node assuming $\SI{0.1}{\mu m}$ depletion thickness and $\SI{2.5}{\fF}$ $C_{par}$. The LET data are retrieved from the NIST Pstar table \cite{pstar}. (b) PW sensing diagram.}
  \label{fig:Vdrop}
\end{figure}

\subsection{Proton detection using a diode}
When a proton interacts with a semiconductor diode, some of the energy deposited in the depletion region of the diode generates electron-hole pairs (EHPs). The average number of EHPs generated in a silicon diode is

\begin{equation} 
EHP = \frac{LET\times t_{dep} \times qu}{\sin\theta_p \times \SI{1.12}{eV}},
\end{equation}
where LET is $dE_{dep}/dx$, $t_{dep}$ is the thickness of the depletion region, $\theta_p$ is the incident angle, and $qu$ is the quenching effect that describes approximately $1/3$ of the deposited energy is used to generate EHPs. The other $2/3$ is either dissipated by heat or via fast recombination of EHPs. The value $\SI{1.12}{eV}$ represents the bandgap energy of the silicon. LET is a highly non-linear function of proton energy. Note that because the proton beam angle from the source is fixed and we know the sensor orientation, mean $\theta_p$ can be easily identified.  

Fig. \ref{fig:diode} depicts the diode sensing mechanism. When the diode is reversely biased by a current source, the generated electrons move to the parasitic capacitance and create a voltage drop of 

\begin{equation}
V_{drop} = \frac{q_e \times EHP}{C_{par}},
\end{equation}
where $q_e$ is the charge of an electron and $C_{par}$ is the parasitic capacitance. Therefore, to achieve single particle sensitivity, a nearly minimum size diode ($\SI{1}{\mu m}$ $\times$ $\SI{1}{\mu m}$) is used to reduce $C_{par}$ because, for a single particle traversing the diode, the average number of EHPs is determined mostly by fabrication parameters and proton energy. In order to have wide detection area, we designed diodes into arrays. When designing an array, we want to maximize the fill factor (defined as the ratio of diode area to the area of the whole circuitry) to capture as many incident particles as possible. Fig. \ref{fig:Vdrop} (a) depicts the average voltage drop by a single proton assuming $C_{par}$ of $\SI{2.5}{fF}$, $t_{dep}$ of $\SI{0.1}{um}$, and a quenching effect of $1/3$. The National Institute of Science and Technology (NIST) pstar table is used to calculate the LET \cite{pstar}. The voltage signal produced during a collision ranges from $\SI{4}{mV}$ to $\SI{78}{mV}$ at $1$$\sim$$67$$~\SI{}{\MeV}$ proton energy range. The voltage drop is a nearly instantaneous event. 

Sensing the instantaneous voltage drop generated during a collision requires high speed analog-to-digital converters which are power-hungry and occupy a large area (especially for an array). In contrast, measuring the time it takes for the generated voltage to return to its baseline is relatively straightforward. We call this delay the pulse width (PW) (See Fig. \ref{fig:Vdrop} (b)). The PW can be expressed as 

\begin{figure}[t]
\centering
  \includegraphics[width=220 pt]{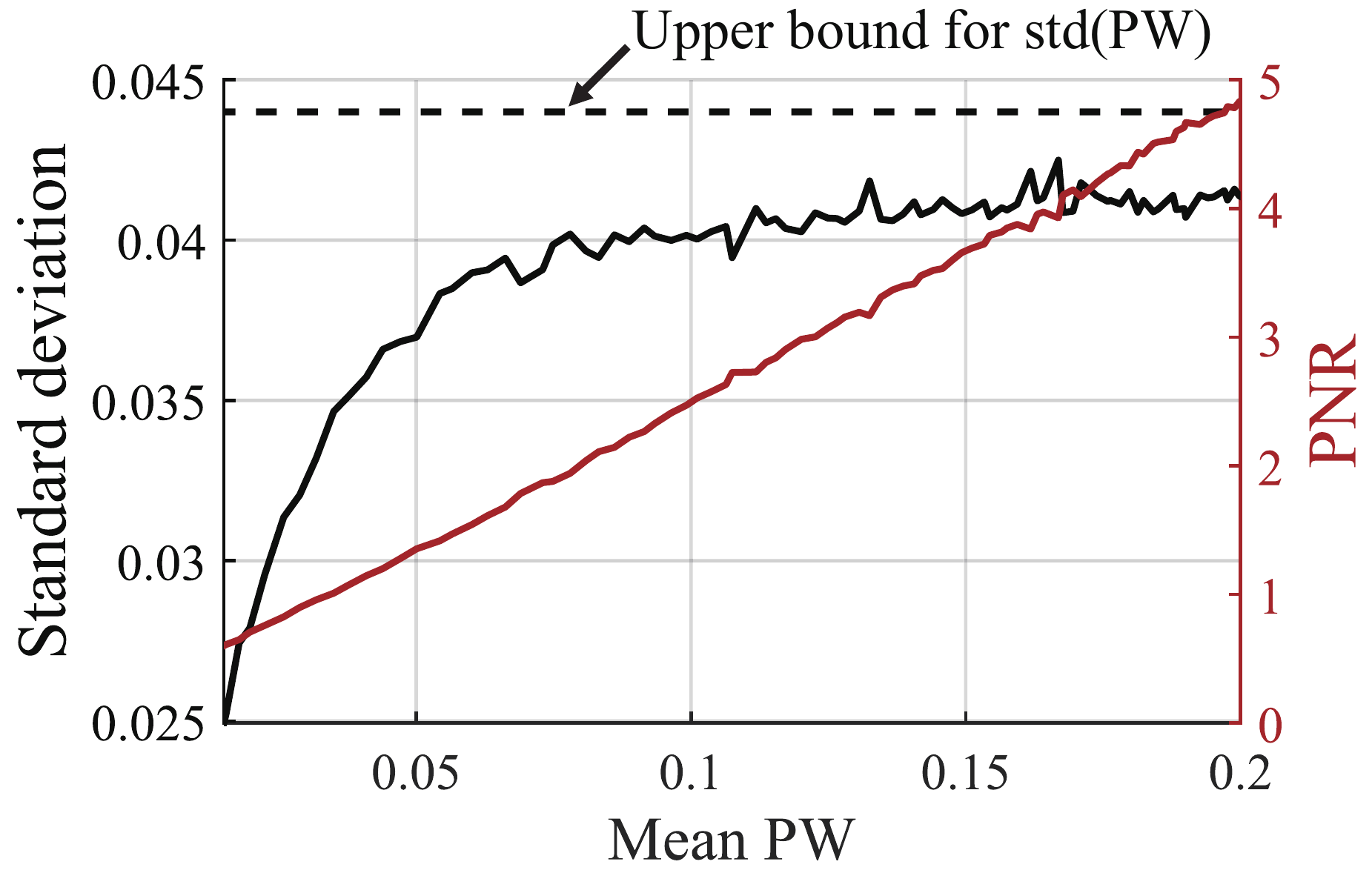}
  \caption{Monte Carlo simulated mean PW versus standard deviation and PNR.}
  \label{fig:PNR}
\end{figure}

\begin{figure*}[!tp]
\centering
  \includegraphics[width=500 pt]{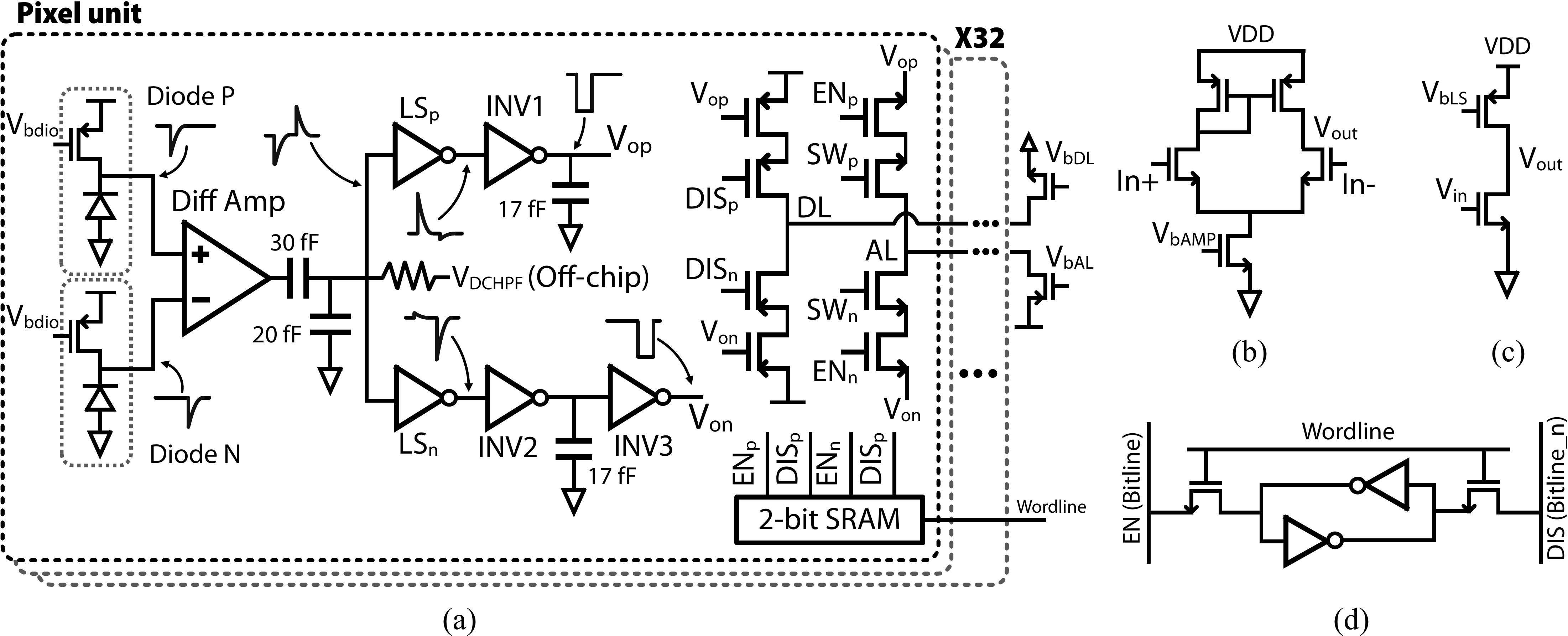}
  \caption{Schematic diagram of (a) pixel unit, (b) differential amplifier, (c) level shifter (LS), and (d) in-pixel 1-bit SRAM.}
  \label{fig:pixel}
\end{figure*}

\begin{align*}
    PW = \tau \ln{\left( \frac{V_{drop}}{V_{th}} \right) } = \tau \ln{\left( \frac{q_e\times qu \times E_{dep}}{1.12 \times \sin\theta_p \times V_{th}\times C_{par}}    \right)},
\end{align*}
where $\tau$ is the time constant at the diode sensing node and $V_{th}$ is the threshold voltage of detection. Even though the sensor output has a logarithmic relationship with $E_{dep}$, this can be pre-calibrated before use.

Including electronic noise at the diode sensing node, $v_{n}$, PW can be expressed as

\begin{equation}
    PW = \tau \ln\left( \frac{V_{drop}}{V_{th}+\overline{v_n}}\right).
\label{eq:log}
\end{equation}
Given this, we can define pulse-width to noise ratio (PNR) as 

\begin{equation}
    \text{PNR} = \frac{\sqrt{\mathbb{E}[PW^2]}}{\sigma_{PW}} > \frac{  \ln(V_{drop}/V_{th})}{\sigma_{v_n}/V_{th}},
\end{equation}
where the delta method is used to find the upper bound for the standard deviation of a logarithmic function. Fig. \ref{fig:PNR} shows the Monte Carlo simulated mean PW versus $\sigma_{PW}$ and PNR. Because PNR is proportional to the mean PW, $V_{drop}$ versus PNR is logarithmic. We can also define signal to noise ratio (SNR) as $V_{drop}/\sigma_{v_n}$. Thus, the ratio of PNR to SNR is 

\begin{equation}
    \frac{\text{PNR}}{\text{SNR}} > \frac{V_{th}\ln(V_{drop}/V_{th})}{V_{drop}},
\end{equation}
which is always less than 1. This means that the PW sensing methodology loses resolution because of the logarithmic transformation of the signal. However, SNR in this analysis assumes perfect sampling of the critical time points of $V_{drop}$ (e.g. the time points corresponding to proton hits) which is impossible in practical situations when using an analog-to-digital converter. The actual PNR loss is subsequently expected to be less than that for the ideal situation.  

\section{System Design}

The design consists of a 64$\times$64 pixel array, a main digital block, a SRAM control block, and a frequency locked loop (FLL). The system must feature low power consumption for future wireless applications, millimeter-scale size, enough detection area with sufficient fill-factor, and robustness to process mismatches.  

The following subsections discuss the analog pixel design, digital system design, FLL, and calibration steps. 

\subsection{Analog Pixel Design}

Fig. \ref{fig:pixel} illustrates the pixel design. To reject common mode noise, a differential sensing scheme is used. Two diodes, diode P and N, are grouped into one pixel unit. A P-type PN diode is used for maximizing the depletion region thickness. A nearly minimum size PMOS current source supplies current to the diode. Changing the bias voltage of the current source ($V_{bdio}$) controls the depletion region thickness, time constant of the sensing node, and DC voltage.

The differential amplifier should feature low input capacitance, high gain, low noise, and low DC output voltage mismatch. Input transistors are critical, as there is a trade-off between the input capacitance and the DC output voltage mismatch. To balance these trade-offs, low $V_{th}$ (LVT) NMOS devices with $\SI{600}{nm}/\SI{600}{nm}$ are used as the input transistors. The differential amplifier occupies a $\SI{4}{\mu m}\times\SI{5.6}{\mu m}$ area.  

A high-pass filter with a $\SI{30}{fF}$ MOMCAP is used to reject the DC output voltage variance of the differential amplifier, and to set the DC voltage to a common voltage uniformly across all pixels by off-chip $V_{DCHPF}$. To provide low $f_{3dB}$, 9 serial pseudo-resistors are used, as the $f_{3dB}$ accuracy is not critical. 

The level shifters ($LS$) shift the DC voltage downward ($LS_{p}$) or upward ($LS_{n}$) to clip signals coming from the other diode. This enables passing signals from the corresponding diode only. The $V_{bLS}$ is an essential variable that controls the trade-off between sensitivity of signal detection ($V_{th}$) and the pixel failure rate (i.e., the ratio between the number of failed pixels and the total number of pixels). For instance, lowering $V_{bLSp}$ increases the output DC voltage of $LS_{p}$, leading to the triggering of the following inverter by a smaller signal. However, it also increases the chance that the noise can trigger the inverter. 

Diode transient pulse is converted to a digital pulse through inverters. The output digital pulses, $V_{op}$ and $V_{on}$, turn on PMOS switches to create the inverted signal on Data Line (DL), which is shared by pixels on the same row.

Due to process, voltage, and temperature (PVT) variations, there is a chance that some pixels are constitutively active and output a false-positive signal even in the absence of radiation events. Because the DL is shared by pixels on the same row, these false positives would hold the DL high and block signals coming from other pixels. Therefore, an in-pixel standard 1-bit 6T SRAM block is implemented to disable any false-positive pixel. Disabling these problematic pixels is called calibration and will be explained in Section \ref{section:cal}.

The overall pixel size is $\SI{8}{\mu m}\times\SI{8}{\mu m}$, leading to a fill-factor of $1/64$. This means that there exists a high chance of protons striking the transistors. Assuming the PN junctions of the transistors have a similar depletion depth to that of the diode, this will create a voltage drop at the node with amplitude 
\begin{equation}
    V_{drop} \leq \SI{250}{mV}\times \frac{C_{par,diode}}{C_{par,circuit}},
\end{equation}
where $C_{par,diode}$ and $C_{par,circuit}$ are the parasitic capacitances at the diode node and the node of the proton hit, respectively. To address this issue, we designed the pixel such that either: 1) the time constant of the node is less than the LSB of the PW sampling, which is $\SI{6}{\mu s}$; or 2) $C_{par,circuit}$ is much greater than $C_{par,diode}$. For instance, the static DC current of the differential amplifier is $\SI{120}{nA}$, and thus the worst case scenario will create a signal
\begin{equation}
PW = \frac{V_{drop}}{dV/dt}=\frac{\SI{250}{mV}\times C_{diode}}{\SI{60}{nA}} \approx \SI{10}{ns}.
\end{equation}
Also, the parasitic capacitance at the HPF is more than 10 times larger than that at the diode node so that the proton hit at the HPF cannot trigger the inverter. 

\begin{figure}[t]
\centering
  \includegraphics[width=252 pt]{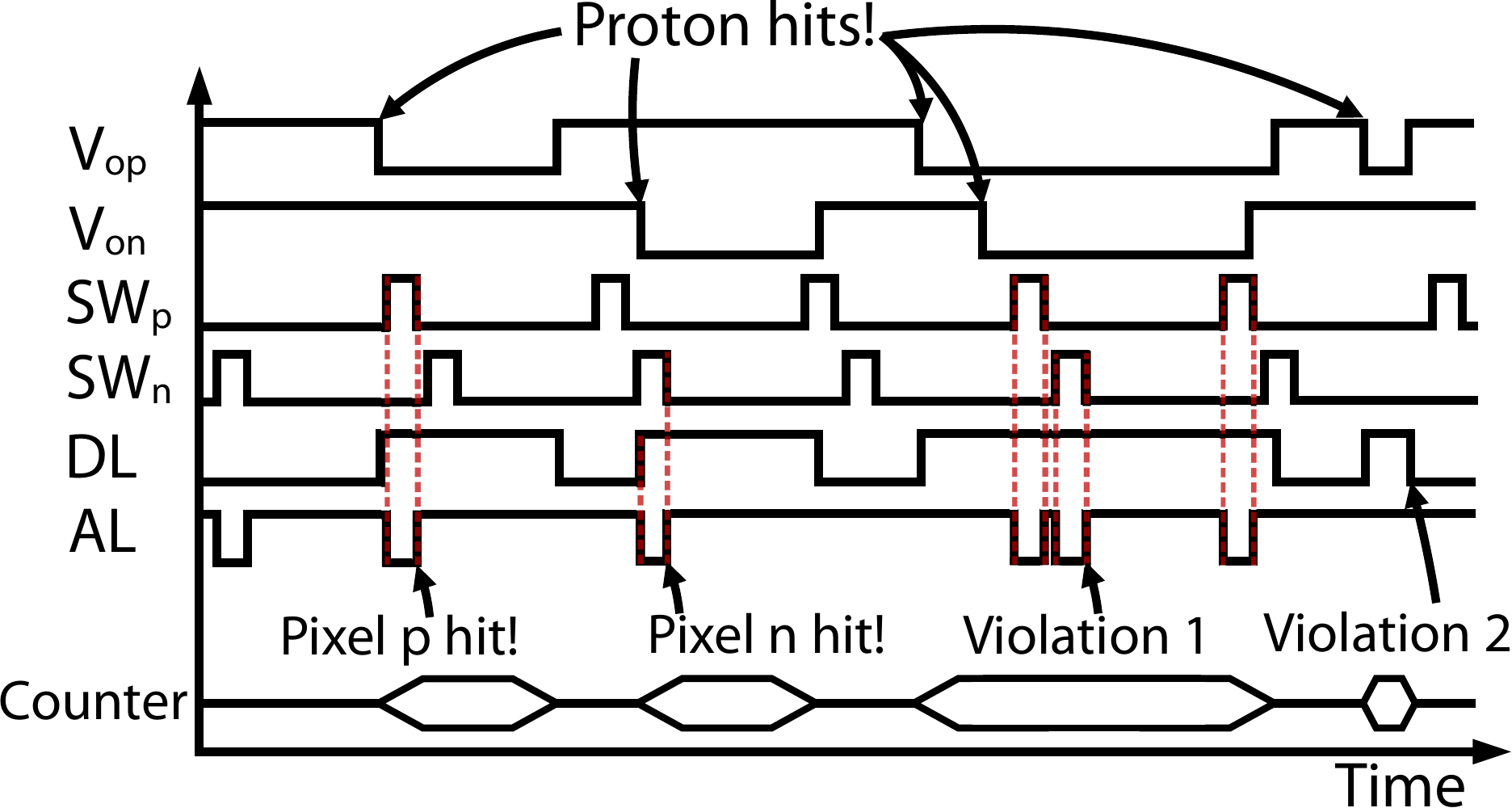}
  \caption{Timing diagram.}
  \label{fig:timing}
\end{figure}

\begin{figure}[!t]
\centering
  \includegraphics[width=250 pt]{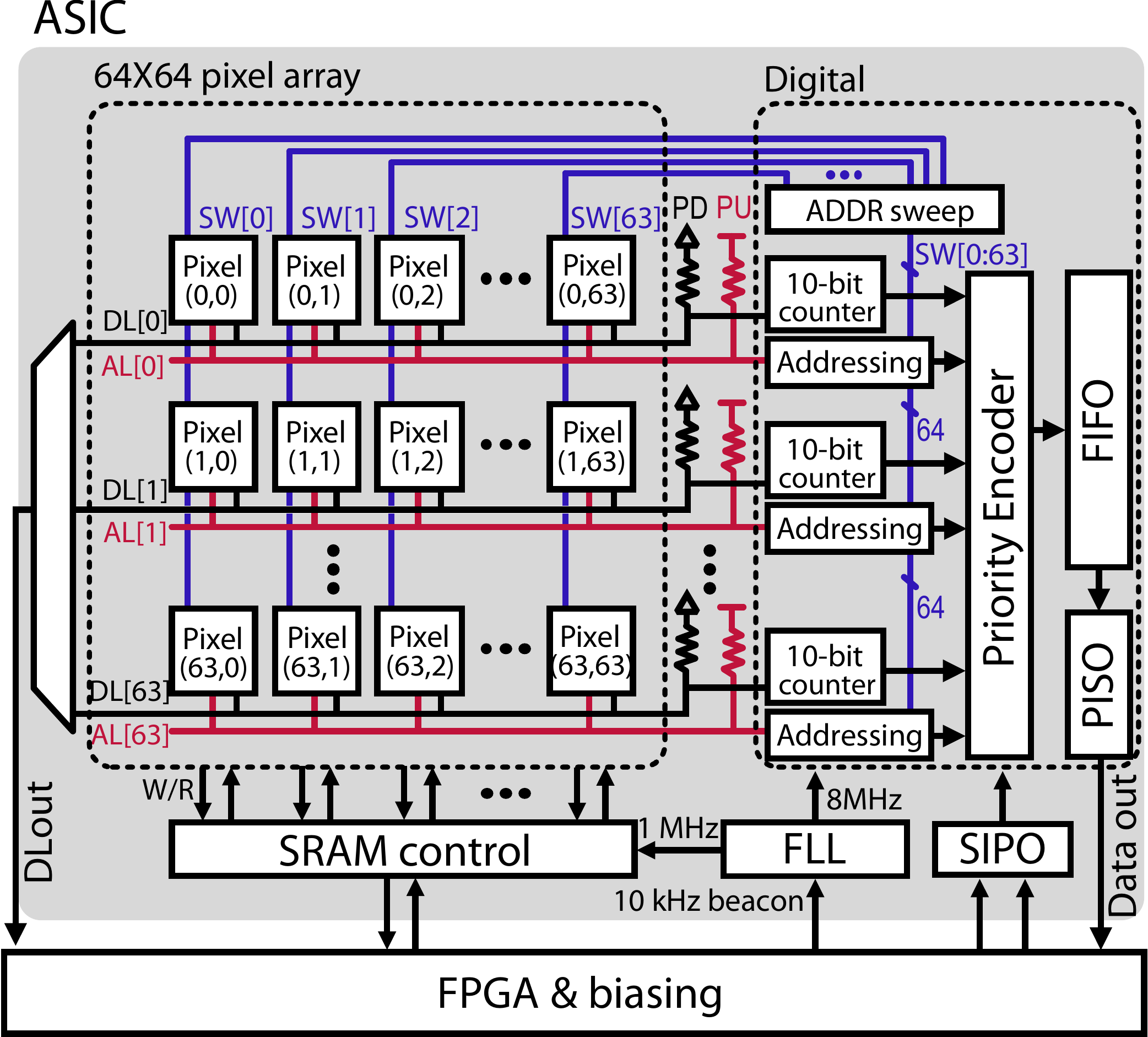}
  \caption{Overall system diagram.}
  \label{fig:digital}
\end{figure}

Fig. \ref{fig:timing} illustrates the timing diagram. A proton hit at the diode creates a voltage pulse at one of the output nodes: $V_{op}$ or $V_{on}$. Then, the same digital pulse but with opposite polarity is created on the DL. The PW is then quantized by a 10-bit digital counter with LSB of $\SI{6}{\mu s}$. Sweeping signals ($SW_p$ and $SW_n$) are 64 non-overlapping periodic $\SI{500}{ns}$ signals generated from the main digital block. The sweeping signals are then transferred to the Address Lines (ALs), which are also shared by the pixels on the same row, when the corresponding pixel has a proton hit and the sweeping pulse exists. Therefore, the main digital block can identify the column address of the proton hit by comparing the sweeping pulse and the AL signal. The sweeping lines are designed so that the overall delay it takes for the signal to travel from the main digital to the pixel and back does not exceed $\SI{500}{ns}$. 

Note that this design methodology cannot distinguish multiple proton hits at different pixels on the same row. This event creates the DL pulse that is the logical OR operation of two voltage output pulses, making the DL pulse inseparable. However, such an event can be easily identified because more than two sweeping pulses will be transferred onto the AL during a single DL signal (See Violation 1 in Fig. \ref{fig:timing}). We can simply discard these events because: 1) this is a rare event; and 2) discarding them will not change the overall statistics because this is a purely stochastic event. Also, the DL PW must be greater than $\SI{500}{ns}\times 64 = \SI{32}{\mu s}$ to guarantee that the column address is accurately identified. If not, there exists a chance that the column address is missing (See Violation 2 in Fig. \ref{fig:timing}). Two types of events, the multiple hit event and the address missing event, are called violation events, and we discard them. The multiple hit violations can be reduced by decreasing the time constant of the diode sensing node. However, this increases the overall power consumption because the 10-bit counter must count faster. We can also mitigate the address missing violations by sweeping the columns faster. Nevertheless, this might result in addressing the wrong columns, as the overall delay of the sweeping signal can exceed its pulse width. Note that the proton hit count can be retrieved even when violation events occur.  

A key advantage of our method is that the pixels consume only static power in the absence of radiation. Unlike traditional imaging applications where every pixel captures the signal periodically, only the struck pixel captures the signal and consumes dynamic power. This is made possible by the PW sensing strategy, and would not be true of a voltage sensing scheme.  
\subsection{Digital System Design}
Fig. \ref{fig:digital} illustrates the overall system. The main digital block features the acquisition of the DL and AL signals, collecting them and buffering, and configuring internal parameters. The SRAM control block manages the enabling and disabling of pixels, as well as the reading of current SRAM values. 

The DL and AL lines on each row have pull-down and pull-up transistors, respectively. The 10-bit counter starts counting at the rising edge of the DL signal. During counting, the addressing block stores the current $SW_{idx}$ value when the AL is high. The counting is finished at the falling edge; and the 10-bit counter value, 6-bit row address, 6-bit column address, and 2-bit status (00 : valid, 01: multiple proton hit, 10 : column address missing) are transferred to the first-in-first-out (FIFO) block.    
    
When multiple rows have data, the priority encoder selects a row that has data and the highest priority. The rows range from 0 to 63, and a lower number translates to higher priority. This prevents data congestion at the interface between the 64 row blocks and the FIFO. The FIFO block has a width of 24 bits and depth of 16. Finally, a parallel-in serial-out (PISO) block receives the data from the FIFO and outputs each data to off-chip.

The maximum latency happens when all rows have data ready and the FIFO is full. Therefore, the best and worst case latency can be expressed as 
\begin{align*}
    t_{latency,min} &\approx PW+t_{cnt}+t_{FIFO}+48t_{PISO} \\
    t_{latency,max} &\approx PW+t_{cnt}+t_{FIFO}+48(16+64)t_{PISO},
\end{align*}
where PW is the pulse width of the data and $t_{cnt}$, $t_{FIFO}$, and $t_{PISO}$ are the clock periods of the counter, FIFO, and PISO, respectively. In default settings, $t_{latency}$ ranges from $PW+\SI{36}{\mu s}$ to $PW+\SI{1932}{\mu s}$. Also, since the PISO block is the bottleneck in transferring data, the maximum proton flux that the digital block can handle is about 41,000 particles per second. 

To give more flexibility of operation, internal parameters can be configured through the serial-in parallel-out (SIPO) from off-chip. $t_{mst}$, $t_{addr}$, and $t_{SRAM}$, which are the periods of the main clock, the addressing clock, and SRAM clock, respectively, as well as $t_{cnt}$ and $t_{PISO}$ can be configured as shown in Table \ref{tab:conf}.   

\begin{table}[]
\caption{\label{tab:conf}Internal Clock Configurations}
\begin{tabular}{@{}lllll@{}}
\toprule
            & Configuration                     & Default Value& Min    & Max  \\ \midrule
$t_{mst}$  & $0.1$, $0.125$, $0.25$, $0.5$, $\SI{1}{\mu s}$  &$\SI{0.125}{\mu s}$& $\SI{0.1}{\mu s}$ & $\SI{1}{\mu s}$\\
$t_{cnt}$  & $t_{mst}\times$$16$, $32$, $48$, $64$, $128$ &$\SI{6}{\mu s}$    & $\SI{1.6}{\mu s}$ & $\SI{128}{\mu s}$\\
$t_{addr}$ & $t_{mst}\times$$2$, $4$, $8$, $32$         &$\SI{0.5}{\mu s}$  & $\SI{0.2}{\mu s}$ & $\SI{32}{\mu s}$ \\
$t_{PISO}$ & $t_{mst}\times$$1$, $2$, $4$, $8$          &$\SI{0.5}{\mu s}$  & $\SI{0.1}{\mu s}$ & $\SI{8}{\mu s}$  \\
$t_{SRAM}$ & $t_{mst}\times$$1$, $8$, $32$, $64$        &$\SI{1}{\mu s}$    & $\SI{0.1}{\mu s}$ & $\SI{64}{\mu s}$     \\ \bottomrule
\end{tabular}
\end{table}

\begin{figure}[!t]
\centering
  \includegraphics[width=250 pt]{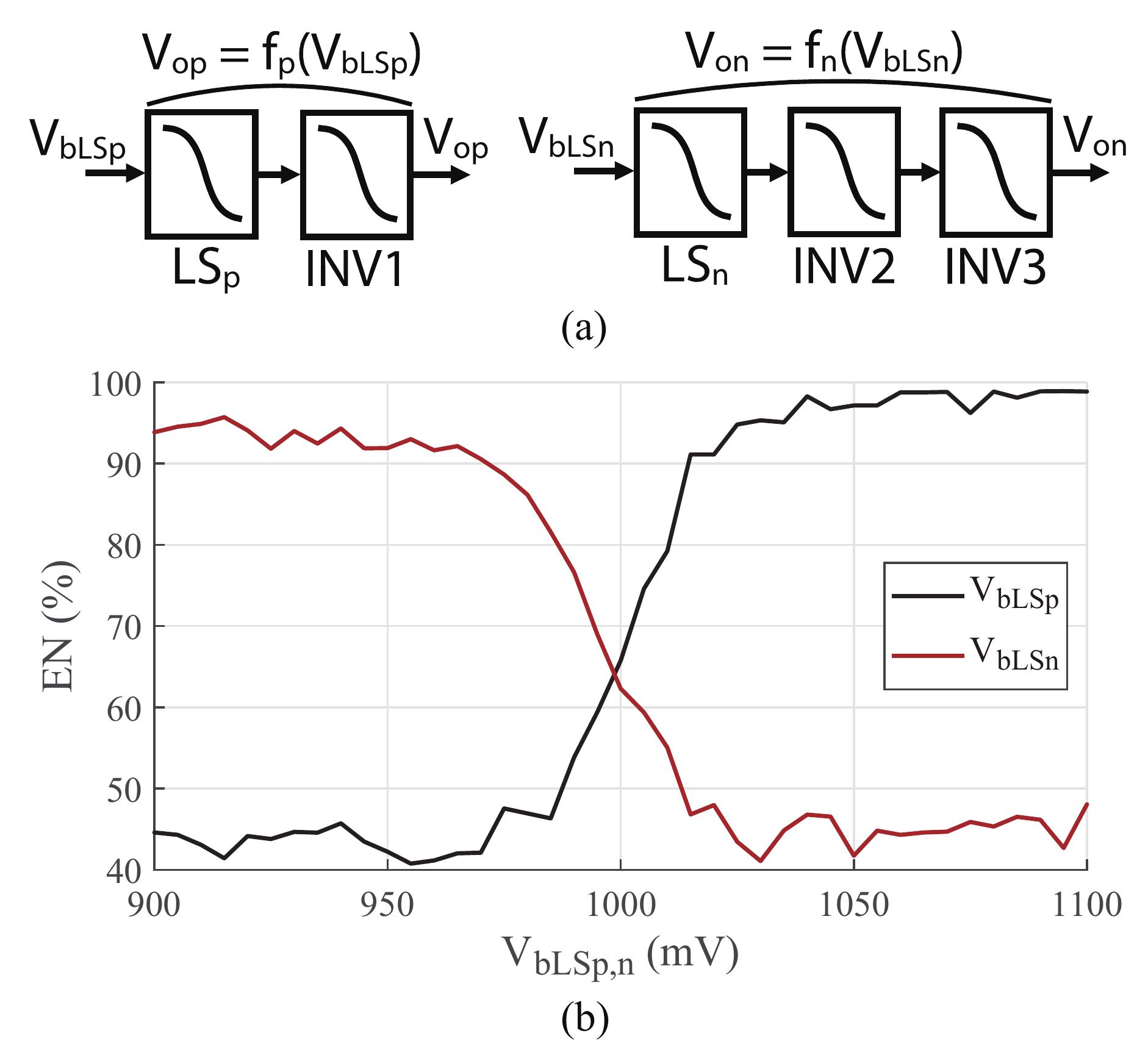}
  \caption{(a) Illustrations of $f_p(V_{bLSp})$ and $f_n(V_{bLSn})$ and (b) measured percentage of enabled pixels versus $V_{bLSp}$ and $V_{bLSn}$. $V_{bLSp,n}$ means $V_{bLSp}$ or $V_{bLSn}$.  }
  \label{fig:cal_sweep}
\end{figure}

\subsection{FLL}
All internal digital clocks are generated from an on-chip FLL. The FLL obviates the need for external bulky crystal oscillators \cite{FLL}. A $\SI{10}{kHz}$ beacon signal is sent from off-chip, and the FLL counts the digitally-controlled oscillator (DCO) clock during each period of the beacon signal. DCO frequency is adjusted through negative feedback based on the difference between the desired number of clocks in one period and the actual counter value. This enables the generation of a $1$$\sim$$\SI{10}{MHz}$ main digital clock with approximately $\SI{280}{kHz}$ frequency resolution.

\subsection{Calibration}
\label{section:cal}
The aforementioned constitutively active pixels are disabled before radiation through calibration steps. Each DL can be monitored off-chip through a 64 to 1 multiplexer. The calibration steps are: 1) disable every pixel through the in-pixel SRAM; 2) enable one pixel and monitor the corresponding DL signal for $\SI{50}{ms}$; 3) disable the pixel if the DL signal is noisy or high; and 4) repeat this process for the remaining pixels. This calibration process is carried out by an external FPGA, and takes approximately 5 minutes. 

By using this technique, we can indirectly measure the statistics of mismatch among pixels. Fig. \ref{fig:cal_sweep}(a) depicts the function $f_p(V_{bLSp})$ and $f_n(V_{bLSn})$. The percentage of enabled pixels after the calibration, $EN$, will be varied based on $V_{bLSp,n}$ value. EN is essentially the percentage of pixels whose $V_{op,n}$ is VDD, which can be described as
\begin{align*}
    \text{EN} \coloneqq 100 \sum_{i=1}^{N}1\{V_{op,n}=VDD\}/N = \mathbb{E}[1\{V_{op,n}=VDD\}],
\end{align*}
where N is the total number of pixels and $V_{op,n}$ means $V_{op}$ or $V_{on}$. Fig. \ref{fig:cal_sweep}(b) shows the measured EN after the calibration when $V_{bLSp}$ and $V_{bLSn}$ are swept from $\SI{900}{mV}$ to $\SI{1100}{mV}$. The slope from $\SI{975}{mV}$ to $\SI{1025}{mV}$ represents the mismatch of the function $V_{op,n} = f_{p,n}(V_{bLSp,n})$ among the pixels. For instance, if the pixels were identical without any mismatch, the slope would be infinite because every pixel becomes enabled at a certain $V_{bLSp,n}$ value; that is, the graph shows the measured cumulative distribution function of $f_{p,n}$. First-order Gaussian fitting of the derivatives of the graphs gives the means of $\SI{1000}{mV}$ and $\SI{995}{mV}$ and the standard deviations of $\SI{19.33}{mV}$ and $\SI{21.85}{mV}$ for $f_{p}(V_{bLSp})$ and $f_{n}(V_{bLSn})$, respectively.   

\begin{figure}[!b]
\centering
  \includegraphics[width=220 pt]{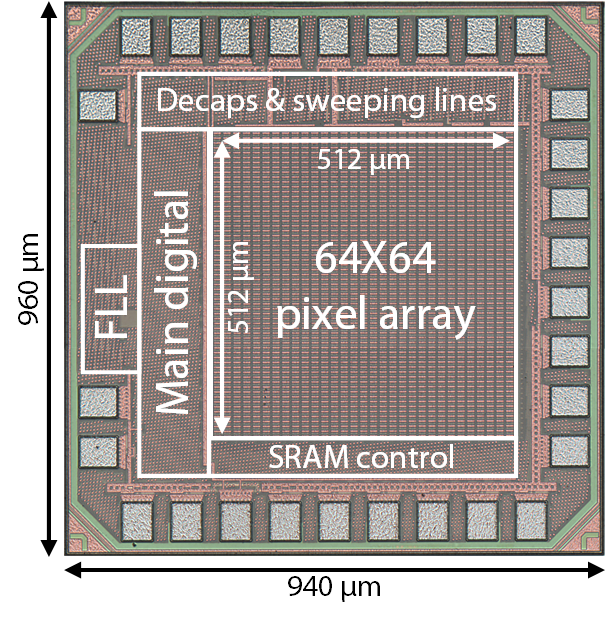}
  \caption{Chip die photo.}
  \label{fig:photo}
\end{figure}

\begin{figure}[!t]
\centering
  \includegraphics[width=220 pt]{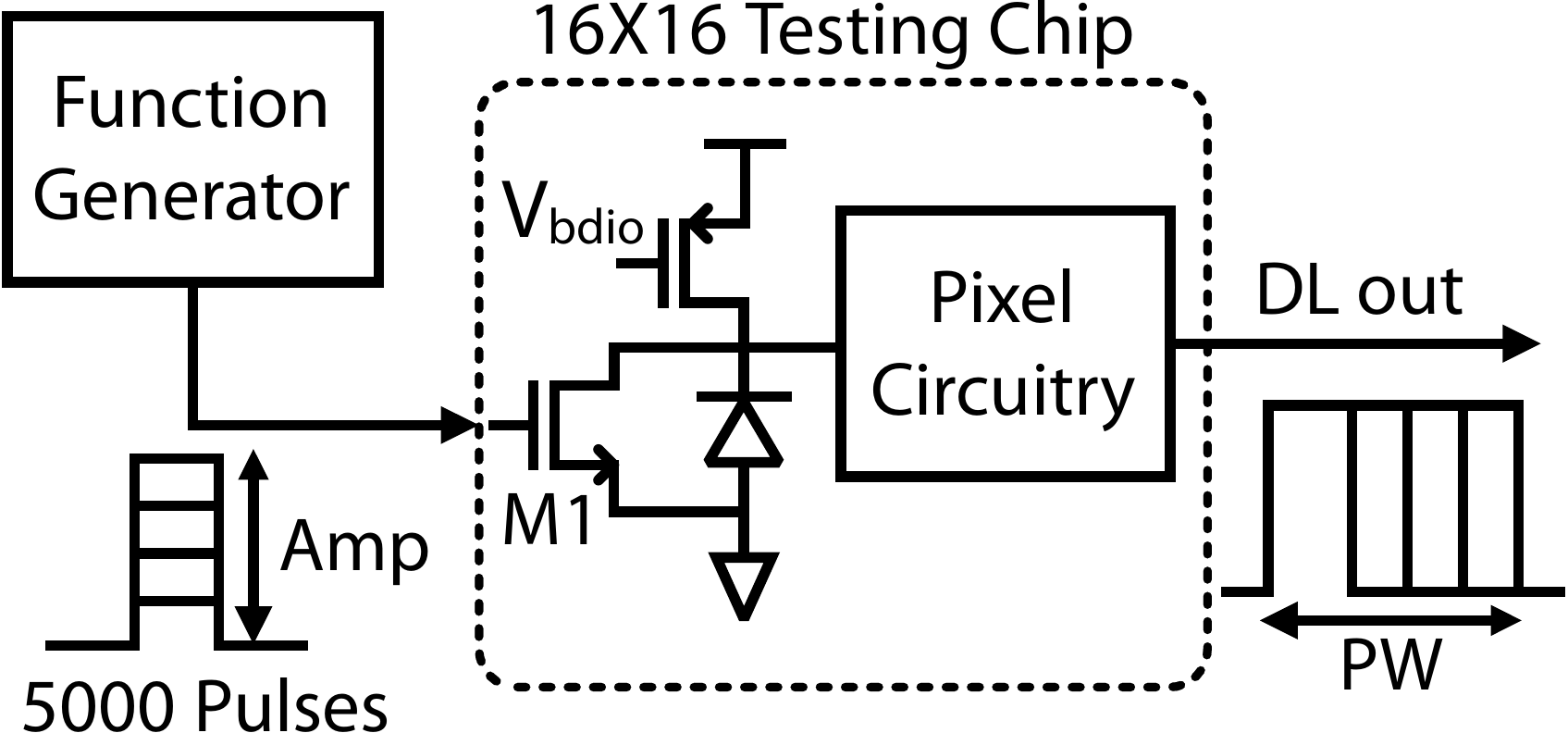}
  \caption{Electrical measurement setup diagram.}
  \label{fig:Etest}
\end{figure}

\section{Measurement Results}
A prototype single charged particle dosimeter system was fabricated in TSMC $\SI{65}{nm}$ Low-power CMOS technology. The ASIC is $\SI{940}{\mu m}$$\times$$\SI{960}{\mu m}$ and its die photo is shown in Fig. \ref{fig:photo}. The detection area is $\SI{512}{\mu m}$$\times$$\SI{512}{\mu m}$ with fill factor of $1/64$.   

This section describes the measurement setups and results. To analyze the electrical noise and pixel-to-pixel variations, a separate 16$\times$16 testing chip was measured. The whole system was verified under a $\SI{67.5}{MeV}$ proton beam generated by a 76-inch cyclotron. The measurement results were compared to those of the Monte Carlo simulation results. 

\begin{figure}[!t]
\centering
  \includegraphics[width=252 pt]{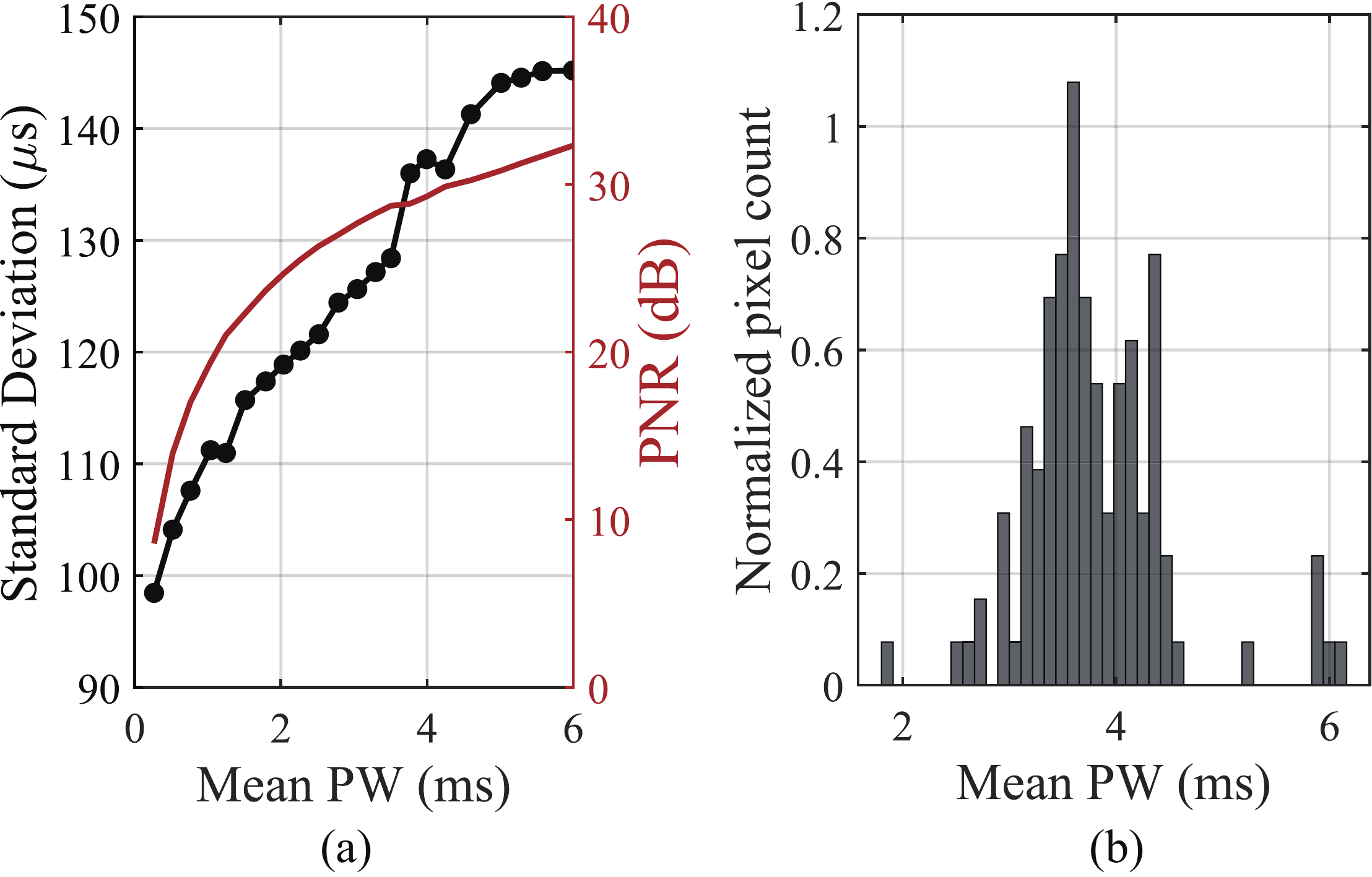}
  \caption{Electrical measurement result of (a) single pixel noise and (b) pixel-to-pixel variation.}
  \label{fig:Enoise}
\end{figure}

\subsection{Electrical Measurement Results}
Fig. \ref{fig:Etest} depicts the electrical measurement setup. A 16$\times$16 testing chip with identical pixel design but with an electrical input was used to characterize noise and pixel-to-pixel variations. The M1 is a nearly minimum size transistor, $\SI{200}{nm}/\SI{60}{nm}$. The function generator generates 5,000 identical pulses with $\SI{1}{\mu s}$ pulse width. These pulses cause an instantaneous voltage drop at the diode node; the rest of the pixel circuitry outputs digital pulses on the DL. The mean and standard deviation of the PW were then measured. These measurements were repeated for various amplitude values of the input pulses. 

Fig. \ref{fig:Enoise}(a) shows the single pixel measurement results. The PW variation increases with the mean PW, and the standard deviation of the PW signals is less than $\SI{150}{\mu s}$ throughout the whole operating range (0-\SI{6}{ms}). PNR is 27.7 dB at $\SI{3}{ms}$ PW.

To measure pixel-to-pixel variation, 5,000 identical pulses were presented to pixels and mean PWs were measured. Fig. \ref{fig:Etest}(b) is the normalized histogram of mean PWs of the 256 pixels. This histogram is essentially the input-to-output PW gain mismatch among the pixels. The variation is mainly due to the mismatch in the differential amplifier and the LS. Note that because the input transistor (M1) mismatch, which is expected to be significant due to small size, is embedded in the measurement result, the actual pixel-to-pixel variation would be smaller. In applications where high-spatial resolution dose map is required, this variation can be pre-calibrated before the treatment by measuring each pixel's responses at different proton energies.  

\begin{figure}[!t]
\centering
  \includegraphics[width=250 pt]{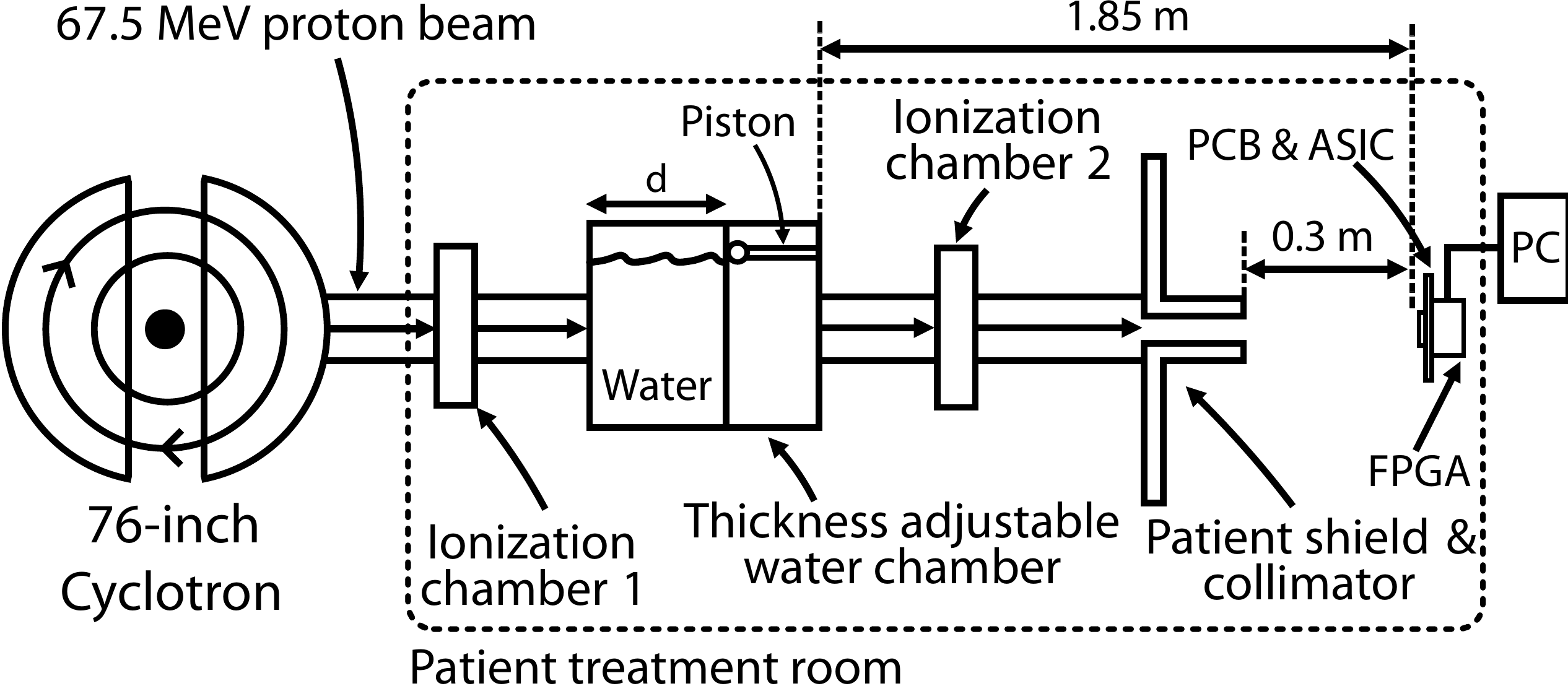}
  \caption{Proton measurement setup diagram at CNL.}
  \label{fig:protonsetup}
\end{figure}

\begin{figure}[!t]
\centering
  \includegraphics[width=220 pt]{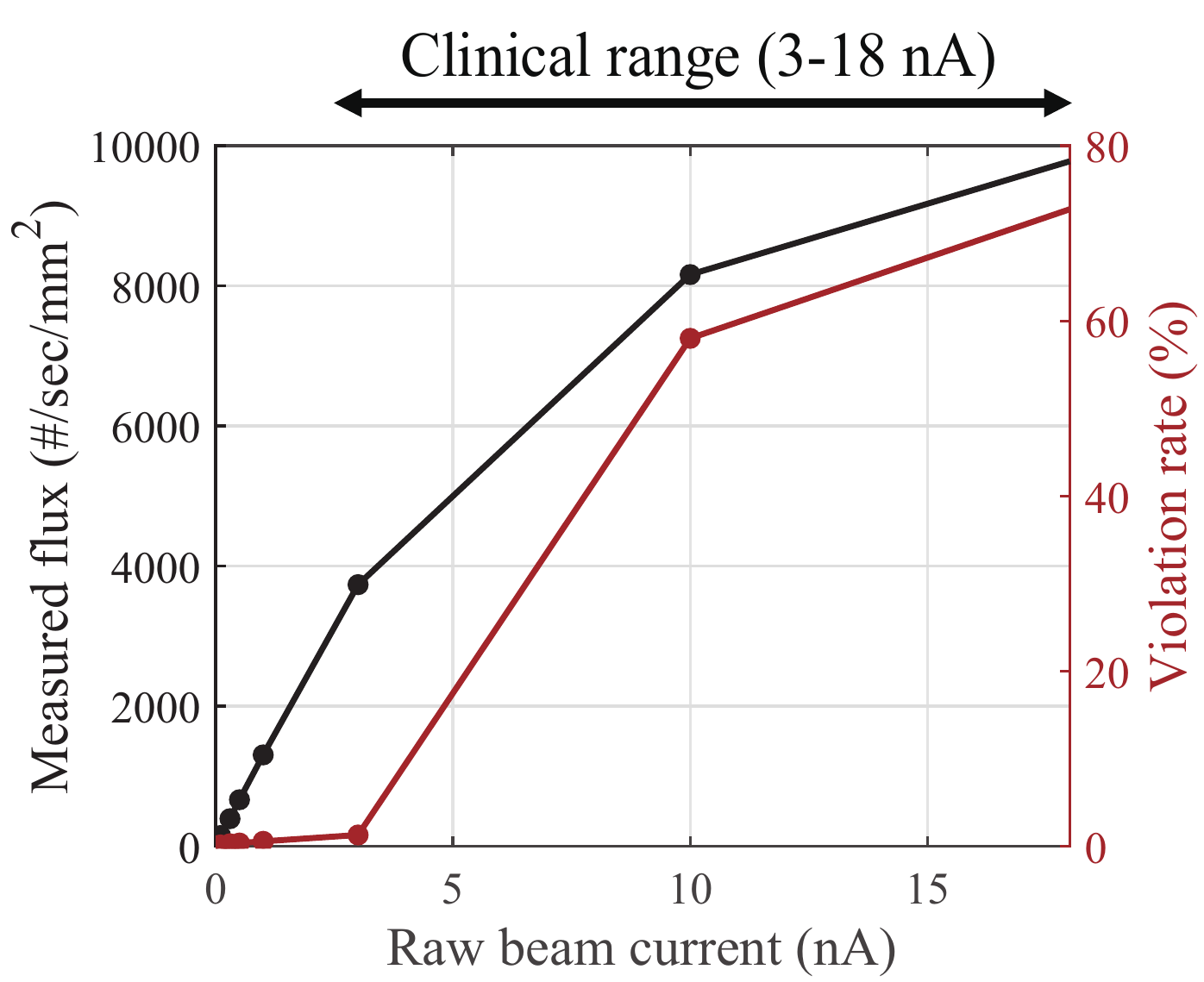}
  \caption{Proton beam current versus measured flux and violation rate.}
  \label{fig:beamcurrent}
\end{figure}

\begin{table}[!b]
\renewcommand{\arraystretch}{1.4}
%\caption{\centering\small{Relationship between the water thickness (d in Fig. \ref{fig:protonsetup}) and the beam characteristics.}}
\centering
\caption{\small{Relationship between the water thickness (d in Fig. \ref{fig:protonsetup}) and the beam characteristics.}}
\centering\label{tab:relation}
\begin{tabular}{lll}
\toprule
\textbf{Water Thickness (d)} & \textbf{Increase}   & \textbf{Decrease}  \\
\midrule
Proton energy loss in the water chamber     & Increase & Decrease \\
Proton energy after the water chamber       & Decrease & Increase \\
Proton energy deposition in the detector~~~~~~& Increase & Decrease \\
Proton flux at the detector                 & Decrease & Increase \\
\bottomrule
\end{tabular}
\end{table}

\begin{figure}[!t]
\centering
  \includegraphics[width=250 pt]{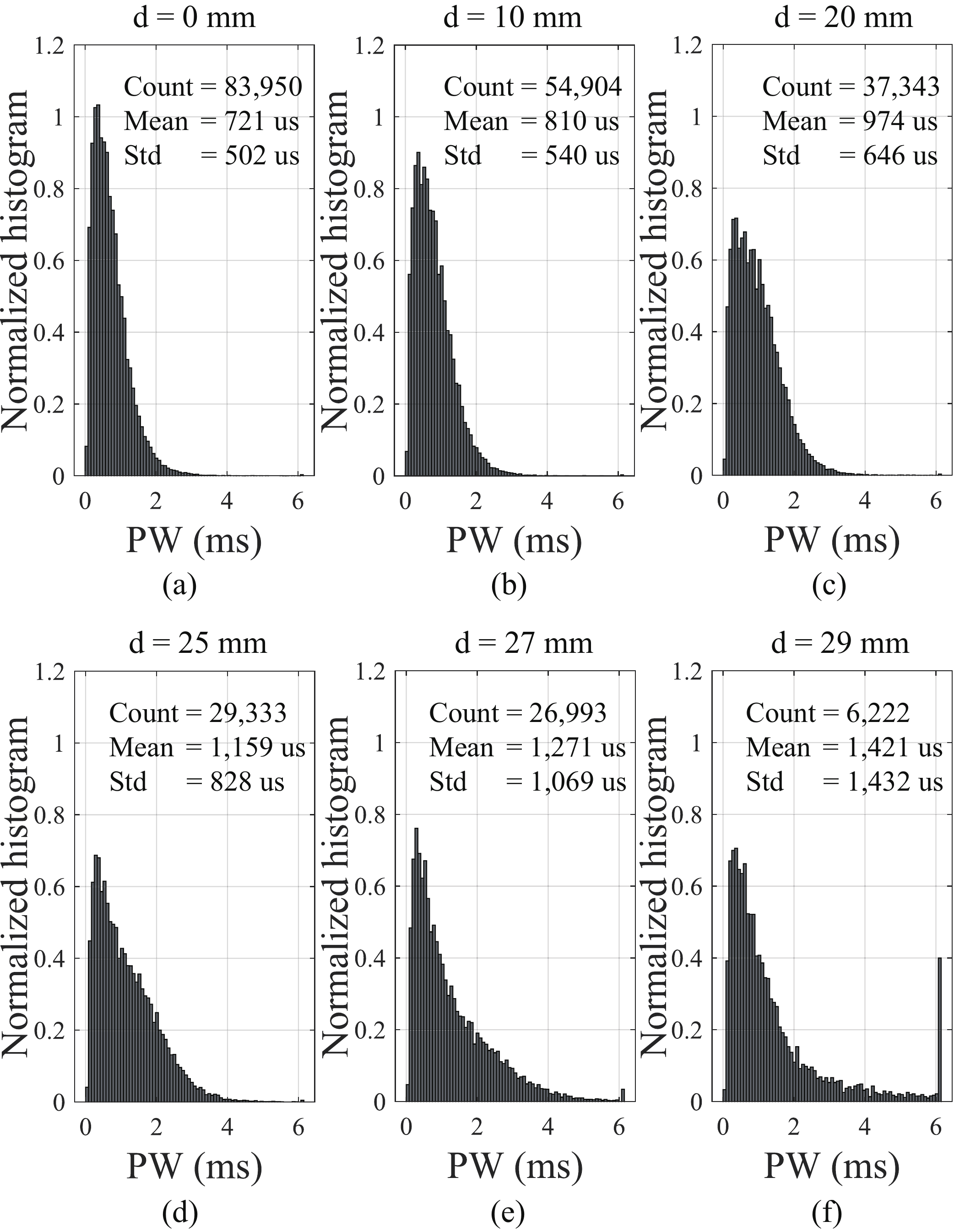}
  \caption{Normalized PW histograms measured for 80 seconds at (a) d = $\SI{0}{mm}$, (b) d = $\SI{10}{mm}$, (c) d = $\SI{20}{mm}$, (d) d = $\SI{25}{mm}$, (e) d = $\SI{27}{mm}$, and (f) d = $\SI{29}{mm}$.}
  \label{fig:pdf}
\end{figure}

\begin{figure*}[!t]
\centering
  \includegraphics[width=475 pt]{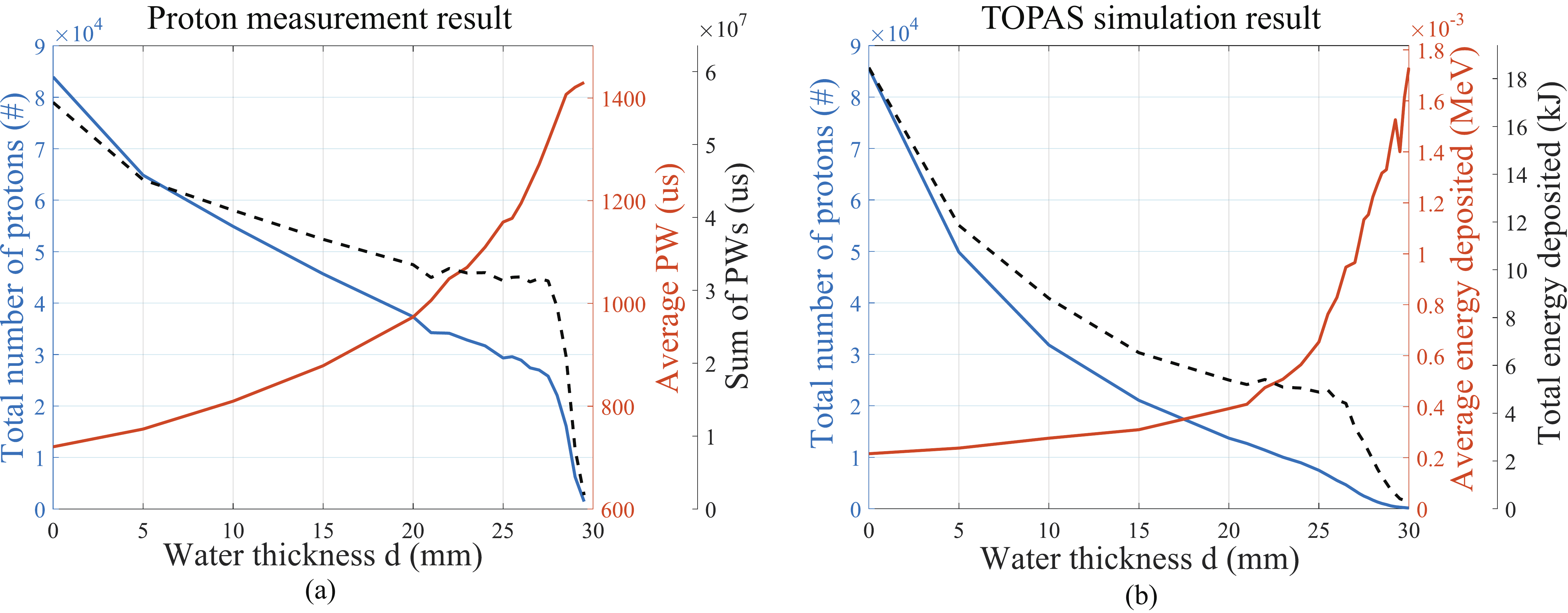}
  \caption{(a) Measured total number of protons, average PW, and summation of PW for 80 seconds of beam time (b) TOPAS simulated total number of protons, average energy deposition, and total energy deposition.}
  \label{fig:meastopas}
\end{figure*}

\subsection{Proton Simulation and Measurement Results}
The prototype ASIC was tested at Crocker Nuclear Laboratory (CNL) at University of California, Davis. The proton beam facility has treated more than 1,700 ocular patients with malignant and benign ocular tumors since 1994 \cite{Inder, Inder2, Bruce}. 

Fig. \ref{fig:protonsetup} depicts a simplified diagram of the proton measurement setup in an eye-treatment room. A $\SI{67.5}{MeV}$ with $\SI{1.3}{MeV}$ full width half maximum (FWHM) proton beam is generated by the 76-inch cyclotron. The beam enters the treatment room and passes through: ionization chamber 1 which monitors the dose; a thickness adjustable water chamber that attenuates the proton beam energy; ionization chamber 2; a patient shield; and lastly a collimator. The ASIC was placed at the position of the patient's eye during treatment (the iso-center) and the data was collected via an FPGA. The beam energy at the patient was controlled by the water chamber. Due to the nature of the energy loss mechanism of the charged particle (Eq. (\ref{eq:bethe})), the energy deposition by protons has an inverse relationship with the proton energy above $\sim$$\SI{0.1}{MeV}$, thus EHP increases with increasing water thickness. Also, as the water thickness increases, the beam scatters more and thus fewer protons reach the detector. This leads to a smaller proton flux, which is the number of protons in unit area per second, at the detector. These relationships are summarized in Table \ref{tab:relation}. 

\begin{table*}[!th]
\renewcommand{\arraystretch}{1.2}
\centering
\caption{\label{tab:compare} Comparison table with related state-of-the-art works.}
\begin{tabular}{llllll}
\toprule
              &\cite{MOS1}&\cite{MOS2}&\cite{RL} &\cite{FG}  & \textit{This work} \\
\midrule
    
Sensing method& Vth shift & Vth shift & RL/OSL   & Floating gate & Diode              \\

Sensing area (mm$^2$ or mm$^3$)     & 0.3$\times$0.05  & 0.2$\times$0.2 & 0.5$\times$0.5$\times$2 (single rod) & 0.1$\times$0.08  & 0.512$\times$0.512        \\

Power consumption for sensing (mW)& N/A (Passive) & N/A (Passive) & 4 (low), 16 (peak) &2& 0.535 \\

Power supply (V)  & N/A & N/A & N/A & 1.2 & 1.2                \\

Real time?        & No  & Yes & Yes & Yes & Yes                \\

\textbf{Single particle detection?} & \textbf{No} & \textbf{No}  & \textbf{No}  & \textbf{No} & \textbf{Yes}\\
\bottomrule
\end{tabular}
\end{table*}

Fig. \ref{fig:beamcurrent} shows measured proton flux and violation rate at different proton beam current settings. The raw beam current is proportional to the actual proton flux. The measured flux increases linearly from $\SI{0.1}{nA}$ to $\SI{3}{nA}$, and starts to saturate after $\SI{3}{nA}$. %The percentage of violated data (violation rate), increases after $\SI{3}{nA}$ because the chance of having multiple hits at different pixels on the same row at the same time increases as proton flux increases. 
Even though the clinical range at CNL is from $\SI{3}{nA}$ to $\SI{18}{nA}$, the remaining inviolate data provides enough data to extract meaningful statistics of energy deposition. We can also decrease the time constant at the diode sensing node to reduce the chance of having violations.    

Normalized PW histograms measured for 80 seconds of the beam time at different water thicknesses are shown in Fig. \ref{fig:pdf}. The 10-bit counter quantized the PW of the DL signals from 0-$\SI{6138}{\mu s}$ with $\SI{6}{\mu s}$ resolution. Any DL signal whose PW is more than $\SI{6138}{\mu s}$ is considered to be saturated. The total proton count decreases as the water thickness increases, mainly due to the proton scattering in the water chamber. As expected, the mean PW, which indirectly measures the mean energy deposition in the depletion region, increases as the water becomes thicker.

The histograms are rightward-skewed and become wider as the water thickness increases. This is mainly due to the Landau effect \cite{Landau}, which is the fluctuation in energy loss by ionization of fast charged particles in a thin layer of matter. This is essentially what the sensor measures: energy loss (PW) by ionization (generation of EHPs) of a charged particle (proton) in a thin layer of matter (depletion region).

\begin{figure}[!t]
\centering
  \includegraphics[width=250 pt]{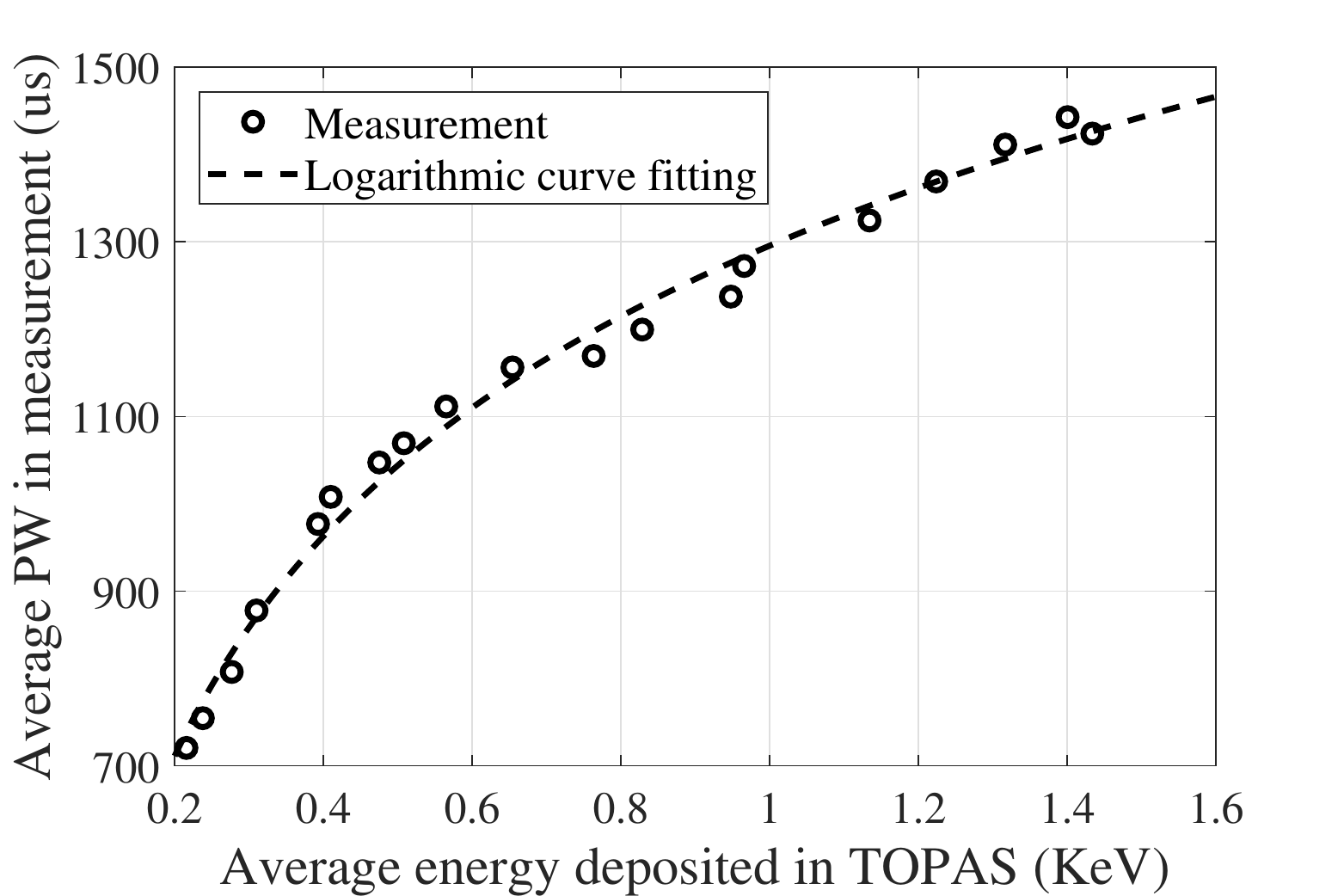}
  \caption{Topas simulated average energy deposition versus measured average PW.}
  \label{fig:curvefit}
\end{figure}

To verify the proton measurement data, the Tool for Particle Simulation (TOPAS) was used \cite{topas,topas1,topas2}. TOPAS wraps and extends the Geometry and Tracking 4 (GEANT4) Monte Carlo particle simulator. GEANT4 is an industry gold-standard for analyzing the behavior of atomic particles \cite{geant4, geant41, geant42}. Fig. \ref{fig:meastopas} (a) shows the measured total number of protons, mean PW, and summation of PW over a 80-second window. The TOPAS simulated total number of protons, mean energy deposition (average energy deposited by each proton), and total energy deposition for all protons, assuming a depletion region thickness of $\SI{0.1}{\mu m}$, is shown in Fig. \ref{fig:meastopas} (b). The trends match well with each other, and we can plot the measured mean PW (orange graph in Fig. \ref{fig:meastopas} (a)) in the $x$-axis and the TOPAS simulated mean energy deposition (orange graph in Fig. \ref{fig:meastopas} (b)) in the $y$-axis to show the energy deposition versus PW relationship (See Fig. \ref{fig:curvefit}). As expected from Eq. \ref{eq:log}, these have a logarithmic relationship. The time constant at the diode sensing node is estimated to be $\SI{363}{\mu s}$ from a curve fit of the data.   

The prototype ASIC consumes average static powers of $\SI{505}{\mu W}$, $\SI{28}{\mu W}$, and $\SI{2}{\mu W}$ for the analog pixel array, digital system, and FLL, respectively. A comparison table with state-of-the-art dosimeters are summarized in Table \ref{tab:compare}. This work has the second largest sensing area of $0.512$$\times$$\SI{0.512}{\mm^2}$, the lowest power consumption among active sensors, and the capability of detecting radiation in real time. Most importantly, this work is the first work that can detect energy deposition by single charged particles with a form-factor and the power consumption compatible with wireless \emph{in-vivo} dosimeter for cancer therapy. This work enables not only the detection of the Bragg peak, but an analysis of the radiation dose's true biological effect.

\section{Conclusion}
A new CMOS diode based 64$\times$64 single charged particle radiation detector is proposed and verified using a clinical proton beam. The design incorporates an analog pixel array with nearly minimum sized diodes, a digital system, SRAM control block, and FLL. Theoretical analysis of measuring charged particles using a diode is presented. The prototype is about $1$$\times$$\SI{1}{mm^2}$ with a detection area of $0.512$$\times$$\SI{0.512}{mm^2}$. The proton measurement results are compared with detailed simulation results. We envision that the proposed system can be used for various cancer therapies, including targeted radionuclide therapy or hadron beam therapy.  

\section*{Acknowledgment}
The authors would like to thank the Chan Zuckerburg Biohub, the Department of Radiation Oncology at UCSF, Crocker Nuclear Laboratory (CNL), Berkeley Sensors and Actuators Center (BSAC), Berkeley Wireless Research Center (BWRC), Sang Min Han of University of California, Berkeley for helpful discussions, and Hyun Joo Song of University of California, Berkeley for her help in the preparation of this manuscript.

\ifCLASSOPTIONcaptionsoff
  \newpage
\fi


\begin{thebibliography}{1}

\bibitem{protonphysics}
W.~Newhauser et al., ``The Physics of Proton Therapy," \emph{Physics in Medicine and Biology}, vol. 60, pp. R155-R209, 2015.

\bibitem{stoi}
M.~Yang et al., ``Comprehensive analysis of proton range uncertainties related to patient stopping-power-ratio estimation using the stoichiometric calibration," \emph{Physics in Medicine and Biology}, vol. 57, pp. 4095-4115, 2012.

\bibitem{MOS1} 
G.~Beyer et al., ``An Implantable MOSFET Dosimeter for the Measurement of Radiation Dose in Tissue During Cancer Therapy," \emph{IEEE Sensors Journal}, vol. 8, pp. 38-51, 2008. 

\bibitem{MOS2}
E.~Gurp et al., ``In Vivo Dosimetry with a Linear MOSFET Array to Evaluate the Urethra Dose During Permanent Implant Brachytherapy Using Iodine-125," \emph{International Journal of Radiation Oncology Biology Physics}, vol. 75, no. 4, pp.1266-1272, 2009. 

\bibitem{RL} 
C.~Andersen et al. ``Characterization of a Fiber-coupled Al$_2$O$_3$:C Luminescence Dosimetry System for Online In Vivo Dose Verification during $^{192}$Ir Brachytherapy," \emph{Medical Physics}, vol. 36, pp. 708-718, 2009. 

\bibitem{FG}
A.~Shamim et al., ``Wireless Dosimeter: System-on-Chip Versus System-in-Package for Biomedical and Space Applications," \emph{IEEE Transactions on Circuits and Systems II}, vol. 55, pp. 643-647, 2008. 

\bibitem{Diode} 
E.~Grusell et al., ``General Characteristics of the Use of Silicon Diode Detectors for Clinical Dosimetry in Proton Beams," \emph{Physics in Medicine and Biology}, vol. 45, pp. 2573-2582, 2000. 

\bibitem{PSD}
L.~Wang et al., ``Determination of the Quenching Correction Factors for Plastic Scintillation Detectors in Therapeutic High-energy Proton Beams," \emph{Physics in Medicine and Biology}, vol. 57, pp.7767-7781, 2012. 

\bibitem{BioEffect}
D.T.~Goodhead et al., ``Mutation and Inactivation of Cultured Mammalian Cells Exposed to Beams of Accelerated Heavy Ions IV. Biophysical Interpretation," \emph{International Journal of Radiation Biology}, vol. 37(2), pp. 135-167, 1980. 

\bibitem{BioEffect2}
D.T.~Goodhead, ``Initial Events in the Cellular Effects of Ionizing Radiations: Clustered Damage in DNA," \emph{International Journal of Radiation Biology}, vol. 65, pp. 7-17, 1994. 

\bibitem{Book} 
E.J.~Hall, ``Radiobiology for the Radiologist, 7th Edition" \emph{Hagerstown, Md. :Medical Dept., Haper \& Row}, 2012. 

\bibitem{KT}
K.~Lee et al., ``A 64$\times$64 Implantable Real-Time Single-Charged-Particle Radiation Detector for Cancer Therapy," \emph{IEEE International Solid-State Circuits Conference (ISSCC)}, 2020. 

\bibitem{Inder} 
I.K.~Daftari et al., ``An Overview of the Control System for Dose Delivery at the UCSF Dedicated Ocular Proton Beam," \emph{International Journal of Medical Physics}, vol. 5, pp. 242-262, 2016. 

\bibitem{Inder2} 
I.K.~Daftari et al., ``Scintillator-CCD Camera System Light Output Response to Dosimetry Parameters for Proton Beam Range Measurement," \emph{Nuclear Instruments and Methods in Physics Research A}, vol. 686, pp. 7-14, 2012. 

\bibitem{Bruce} 
B.A.~Faddegon et al., ``Experimental Depth Dose Curves of a 67.5 MeV Proton Beam for Benchmarking and Validation of Monte Carlo Simulation," \emph{Medical Physics}, vol. 42, pp. 4199-4210, 2015. 

\bibitem{FLL}
W.~Biederman et al., ``A Fully-Integrated, Miniaturized (\SI{0.125}{mm^2}) \SI{10.5}{uW} Wireless Neural Sensor," \emph{IEEE Journal of Solid-State Circuits}, vol. 48, pp. 960-970, 2013.

\bibitem{Landau}
L.~Landau, ``On the Energy Loss of Fast Particles by Ionization," \emph{Journal of Physics}, vol. 8, 1944. 

\bibitem{topas}
TOPAS : tool for particle simulation : http://topasmc.org

\bibitem{topas1}
J. Perl et al., ``TOPAS - An Innovative Proton Monte Carlo Platform for Research and Clinical Applications," \emph{Medical Physics}, vol. 39, pp. 6818-6837,2012.

\bibitem{topas2}
B. Faddegon el al., ``The TOPAS Tool for Particle Simulation, a Monte Carlo Simulation Tool for Physics, Biology and Clinical Research," \emph{Physica Medica: European Journal of Medical Physics}, vol. 72, pp. 114-121, 2019. 

\bibitem{geant4}
S. Agostinelli et al., ``GEANT4 - A Simulation Toolkit," \emph{Nuclear Instruments and Methods in Physics Research Section A}, vol. 506, pp. 250-303, 2003.

\bibitem{geant41}
J. Allison et al., ``GEANT4 Developments and Applications," \emph{IEEE Transactions on Nuclear Science}, vol. 53, pp. 270-278, 2006.

\bibitem{geant42}
J. Allison et al., ``Recent Developments in GEANT4," \emph{Nuclear Instruments and Methods in Physics Research Section A}, vol. 835, pp. 186-225, 2016.

\bibitem{pstar}
NIST measured proton stopping power data : \\ http://physics.nist.gov/PhysRefData/Star/Text/PSTAR.html

\end{thebibliography}
\end{document}